\documentclass[a4paper, 12pt, openany, oneside]{article}
\usepackage[cp1251]{inputenc} 
\usepackage[english, russian]{babel}
\usepackage{amsmath, latexsym,  amssymb, array, graphics,
amsfonts, amsthm, amsmath,  amsfonts, array, bm, eucal}

\makeatletter
\renewcommand{\@biblabel}[1]{#1.\hfill}
\newcommand{\intl}{\int\limits}

\renewcommand{\Re}{\mathop{\rm Re\,}}

\newcommand{\sign}{\mathop{\rm sign\,}}

\makeatother

\usepackage[dvips]{graphicx}
{

\begin{document}
\renewcommand{\abstractname}{Abstract}
\newcommand{\mc}[1]{\mathcal{#1}}
\newcommand{\E}{\mc{E}}

\thispagestyle{empty} \large

\renewcommand{\refname}{\begin{center} \bf REFERENCES  \end{center}}
\large

\begin{center}
  \bf
PLASMA WAVES REFLECTION FROM A BOUNDARY WITH SPECULAR  ACCOMMODATIVE
BOUNDARY CONDITIONS
\end{center}

\begin{center}
 \;{\bf N. V. Gritsienko, A. V. Latyshev, A. A. Yushkanov}\\
\end{center}

\begin{center}
{\normalsize {\it Moscow State Regional University,
Radio st.,10a, Moscow 105005, Russia}

e-mail: natafmf@yandex.ru, avlatyshev@mail.ru,
yushkanov@inbox.ru}

\end{center}
\date{\today}

\begin{abstract}
In the present work the linearized problem of plasma wave reflection
from a boundary of a half--space is solved analytically. Specular
accommodative conditions of plasma wave reflection from plasma
boundary are taken into consideration. Wave reflectance is found as
function of the given parameters of the problem, and its dependence
on the normal electron momentum accommodation coefficient is shown by
the authors. The case of resonance when the frequency of
self-consistent electric field oscillations is close to the proper
(Langmuir) plasma oscillations frequency, namely, the case of long
wave limit is analyzed in the present paper.
Refs. 17. Figs. 6.\\


\end{abstract}

\begin{center}\bf
1. BASIC EQUATIONS
\end{center}

Research of degenerate electron plasma behaviour, processes which
take place in plasma under electric field, plasma waves becomes more
and more actual at the present time in connection with the problems
of such intensively developing areas as microelectronics and
nanotechnologies  \cite{1} -- \cite{Liboff}.

This work continues the research of the electron plasma behaviour in
external longitudinal alternating electric field \cite{4} --
\cite{8}.

In the present work linearized problem of plasma wave reflection
from a boundary of a half-space of conductive medium is solved
analytically. Specular accommodative conditions of electron
reflection from plasma boun\-dary are taken into consideration. The
diffuse boun\-dary conditions were con\-si\-de\-ra\-ted in
\cite{5}--\cite{7}.

Expression for wave reflectance is obtained and it is shown that in
the case when normal electron momentum accommodation coefficient
takes on a value of zero the wave reflectance is expressed by the
formula obtained earlier in \cite{5}, \cite{8}.

Let us consider degenerate plasma which is situated in a half-space
$x>0$. We assume that self-consistent electric field
$\mathbf{E}(\mathbf{r},t)$ inside plasma has one $x$--component and
varies along the axis $x$ only: $ \mathbf{E}=\{E_x(x,t),0,0\}. $

In this case the electric field is perpendicular to the plasma
boundary which is situated in the plane $x=0$.

Here $\omega_p$ is Langmuir (proper) plasma oscillation frequency,
$\omega_p=\dfrac{4\pi e^2N}{m}$, $N$ is the electron numerical
density (concentration), $m$ is the mass of the electron.

Let us take the system of equations which describes plasma
behaviour. As the kinetic equation we take $\tau$--model Vlasov ---
Boltzmann kinetic equation:
$$
\dfrac{\partial f}{\partial t}+\mathbf{v}\dfrac{\partial f}{\partial
\mathbf{r}}+e\mathbf{E}\dfrac{\partial f}{\partial \mathbf{p}}=
\dfrac{f_{eq}(\mathbf{r},t)-f(\mathbf{r},\mathbf{v},t)}{\tau}.
\eqno{(1.1)}
$$

Here $f=f(\mathbf{r},\mathbf{v},t)$ is the electron distribution
function, $e$ is the electron charge, $\mathbf{p}=m\mathbf{v}$ is
the electron momentum, $m$ is the electron mass, $\tau$ is the
characteristic time period between two collisions,
$f_{eq}=f_{eq}(\mathbf{r},t)$ is the local equilibrium distribution
function of Fermi and Dirac, $ f_{eq}=\Theta(\E_F(t,x)-\E), $ where
$\Theta(x)$ is the function of Heaviside,
$$
\Theta(x)=\left\{\begin{array}{c}
                  1, \quad x>0, \\
                   0,\quad x<0,
                 \end{array}\right.
$$
$\E_F(t,x)=\frac{1}{2}mv_F^2(t,x)$ is the disturbed kinetic energy
of Fermi, $\E=\frac{1}{2}mv^2$ is the kinetic energy of the eletron.

Let us consider the Maxwell equation for the electric field
$$
{\rm div}\,{\mathbf{E}(\mathbf{r},t)}=
4\pi e\int (f(\mathbf{r}, \mathbf{v},t)-f_{0}(\mathbf{v}))
\,d\Omega_F,.
\eqno{(1.2)}
$$
where
$$
d\Omega_F=\dfrac{2d^3p}{(2\pi\hbar)^3}, \qquad
d^3p=dp_xdp_ydp_z.
$$

Here $f_{0}$  is the undisturbed electron distribution function of
Fermi and Dirac, $ f_{0}(\E)=\Theta(\E_F-\E), $ $\hbar$ is the
Planck constant,  $\E_F=\frac{1}{2}mv_F^2$ is the undisturbed
kinetic energy of Fermi, $v_F$ is the electron velocity on the Fermi
surface which is considered as spherical.

Let us search for the solution of the system (1.1) and (1.2) in the
following form
$$
f=f_0(\E)+\E_F \delta(\E_F-\E)
H(x,\mu,t), \qquad \mu=\dfrac{v_x}{v}.
$$

We obtain (see \cite{6}) linearized system of equations of Vlasov
--- Max\-well:
$$
\dfrac{\partial H}{\partial t_1}+\mu \dfrac{\partial H}{\partial
x_1}+ H(x_1,\mu,t_1)=\mu e(x_1,t_1)+\dfrac{1}{2}\int\limits_{-1}^{1}
H(x_1,\mu',t_1)d\mu',
\eqno{(1.3)}
$$
$$
\dfrac{\partial e(x_1,t_1)}{\partial x_1}=\dfrac{3\omega_p^2}
{2\nu^2}\int\limits_{-1}^{1}H(x_1,\mu',t_1)d\mu'.
\eqno{(1.4)}
$$

Here $e(x_1,t_1)$ is dimensionless function
$$
e(x,t)=\dfrac{ev_F}{\nu \E_f}E_x(x,t),
$$
$x_1=x/l$ is the dimensionless coordinate, where $\;l=v_F\tau$ is
the average free path of electrons, $t_1=\nu t$ is the dimensionless
time , $\nu$ is the effective frequency of electron scattering,
$\nu=1/\tau$.

Supposing that $k$ is the dimensional wave number, we introduce the
dimen\-si\-on\-less wave number $k_1=k\dfrac{v_F}{\omega_p}$, then we have
$kx=\dfrac{k_1x_1}{\varepsilon}$, where
$\varepsilon=\dfrac{\nu}{\omega_p}$. Let us introduce the quantity
$\omega_1=\omega\tau=\omega/\nu$.

\begin{center}\bf
2.  BOUNDARY CONDITIONS STATEMENT
\end{center}

Let the plasma wave move to the plasma boundary situated in the
plane $x_1=0$. The electric field of the wave changes according to
the following law
$$
e_+(x_1,t_1)=E_1\exp(-i(\dfrac{k_1x_1}{\varepsilon}+\omega_1
t_1)).
\eqno{(2.1)}
$$
The amplitude of this wave $E_1$ we assume to be given. On the
plasma boundary this wave reflects and the electric field of the
reflected wave has the following form
$$
e_-(x_1,t_1)=E_2\exp(i(\dfrac{k_1x_1}{\varepsilon}-\omega_1
t_1)).
\eqno{(2.2)}
$$

The amplitude $E_2$ is unknown and is to be found from the problem
solution. The quantities $\omega_1$ and $k_1$ are not independent,
the following depen\-den\-ce $ \omega_1=\omega_1(k_1) $ is determined
from the solution of the dispersion equation which will be
introduced below.

It is required to determine what part of the wave energy is absorbed
under the wave reflection from the plasma boundary, and what part of
the energy is reflected, and also to find the phase shift of the
wave. It means we have to calculate the reflectance which is
determined as square of module of the ratio of reflected and
incoming waves amplitudes

$$
R(k,\omega,\varepsilon)=\left|\dfrac{E_2}{E_1}\right|^2
\eqno{(2.3)}
$$
and to find the argument of the amplitudes ratio
$$
\phi(k,\omega,\varepsilon)=\arg\Big(\dfrac{E_2}{E_1}\Big)=\arg E_2-\arg E_1.
\eqno{(2.4)}
$$

Let us outline the time variable of the functions $H(x_1,\mu,t_1)$
and $e(x_1,t_1)$, assuming
$$
H(x_1,\mu,t_1)=e^{-i\omega_1t_1}h(x_1,\mu),\qquad
e(x_1,t_1)=e^{-i\omega_1t_1}e(x_1).
\eqno{(2.5)}
$$

The system of equations (1.3) and (1.4) in this case will be
transformed to the following form:
$$
\mu\dfrac{\partial h}{\partial x_1}+(1-i\omega_1)h(x_1,\mu)=
\mu e(x_1)+\dfrac{1}{2}\int\limits_{-1}^{1}h(x_1,\mu')d\mu',
\eqno{(2.6)}
$$
$$
\dfrac{de(x_1)}{dx_1}=\dfrac{3\omega_p^2}{2\nu^2}
\int\limits_{-1}^{1}h(x_1,\mu')d\mu'.
\eqno{(2.7)}
$$

Further instead of $x_1,t_1$ we write $x,t$. We rewrite the system
of equations (2.6) and (2.7) in the form:
$$
\mu\dfrac{\partial h}{\partial x}+z_0h(x,\mu)=
\mu e(x)+\dfrac{1}{2}\int\limits_{-1}^{1}h(x,\mu')d\mu',\quad
z_0=1-i\omega_1.
\eqno{(2.8)}
$$
$$
\dfrac{de(x)}{dx}=\dfrac{3}{2\varepsilon^2}
\int\limits_{-1}^{1}h(x,\mu')d\mu'.
\eqno{(2.9)}
$$

We consider the external electric field outside the plasma limit is
absent. This means that for the field inside plasma on the plasma
boundary the following condition is satisfied:
$$
e(0)=0.
\eqno{(2.10)}
$$

The non-flowing condition for the particle (electric current) flow through the plasma
boundary means that
$$
\int\limits_{-1}^{1}\mu\,h(0,\mu)\,d\mu=0.
\eqno{(2.11)}
$$

In the kinetic theory for the description of the surface properties
the accommodation coefficients are used often. Tangential momentum
and energy accommodation coefficients are the most--used. For the
problem considered the normal electron momentum accommodation under
the scattering on the surface has the most important significance.

The normal momentum accommodation coefficient
is defined by the follo\-wing relation:
$$
\alpha_p=\dfrac{P_i-P_r}{P_i-P_s}, \quad 0\leqslant \alpha_p
\leqslant 1,
\eqno{(2.12)}
$$
where $P_i$ and $P_r$  are the flows of normal to the surface
momentum of incoming to the boundary and reflected from it electrons,
$$
P_i=\int\limits_{-1}^{0}\mu^2h(0,\mu)d\mu,\qquad
P_r=\int\limits_{0}^{1}\mu^2h(0,\mu)d\mu,
\eqno{(2.13)}
$$
quantity $P_s$ is the normal momentum flow for electrons
reflected from the surface which are in thermodynamic
equilibrium with the wall,

$$
P_s=\int\limits_{0}^{1}\mu^2h_s(\mu)d\mu,\qquad\text{где}\quad
h_s(\mu)=A_s,\quad 0<\mu<1.
\eqno{(2.14)}
$$
The function $h_s(\mu)$ is the equilibrium distribution
function of the correspon\-ding electrons.
This function is to satisfy the condition
similar to the non-flowing condition:
$$
\int\limits_{-1}^{0}\mu h(0,\mu)d\mu+\int\limits_{0}^{1}\mu
h_s(\mu)d\mu=0.
\eqno{(2.15)}
$$

We are going to consider the relation between the normal momentum
accom\-mo\-da\-tion coefficient $\alpha_p$ and the diffuseness coefficient
$q$ for the case of specu\-lar and diffuse boundary conditions which
are written in the following form:
\[
h(0,\mu)=(1-q) h(0,-\mu)+a_s, \quad 0<\mu<1.
\]

Here $q$ is the diffuseness coefficient ($0\leqslant q\leqslant 1$),
$a_s$ is the quantity determined from the non-flowing condition.

From the non-flowing condition we derive
\[
a_s=-2q\int\limits_{-1}^{0}\mu h(0,\mu)d\mu=qA_s.
\]

After that we find the difference between the flows
\[
P_i-P_r=q\int\limits_{-1}^{0}\mu^2 h(0,\mu)d\mu-\int\limits_{0}^{1}\mu^2
a_sd\mu=
\]
\[
=q\int\limits_{-1}^{0}\mu^2 h(0,\mu)d\mu-q\int\limits_{0}^{1}\mu^2
A_sd\mu=qP_i-qP_s.
\]
Substituting the expressions obtained to the definition of
the normal momen\-tum accommodation coefficient we obtain that
$\alpha_p=q$.

Thus, for the specular and diffuse boundary conditions
the normal momen\-tum accommodation coefficient
$\alpha_p$ coincides with the diffusion
coefficient $q$.

Together with the specular and diffuse boundary conditions
other variants of boundary conditions are used in the kinetic
theory.

In particular, accommodative boundary conditions are widely used.
They are divided into two forms: diffuse accommodative and specular
accom\-mo\-dati\-ve boundary conditions (see \cite{9}).

We consider specular accommodative boundary conditions.
For the func\-tion $h$ these conditions
will be written in the following form:
$$
h(0,\mu)=h(0,-\mu)+A_1+A_2\mu, \quad 0<\mu<1.
\eqno{(2.16)}
$$

Coefficients $A_1$ and $A_2$ can be derived from the non-flowing
condition and the definition of the normal electron momentum
accommodation coefficient.

The problem statement is completed. Now the problem consists in
finding of such solution of the system of equations (2.8) and (2.9),
which saatisfies the boundary conditions (2.10)--(2.16). Further,
with use of the amplitudes of reflected and incoming waves found it
is required to find the reflectance of the incoming wave energy
(2.3) and the argument of the amplitudes ratio (2.4).\\

\begin{center}\bf
3. THE RELATION BETWEEN FLOWS
AND BOUNDARY CONDITIONS
\end{center}

First of all let us find expression which relates the constants
$A_0,\;A_1$ from the boundary condition (2.13).

To carry this out we will use the condition of non-flowing
(2.12) of the particle flow through the plasma boundary, which
we will write as a sum of two flows:
$$
N_0\equiv \int\limits_{0}^{1}\mu h(0,\mu)d\mu+\int\limits_{-1}^{0}
\mu h(0,\mu)d\mu=0.
$$

After evident substitution of the variable in the second
integral we obtain:
$$
N_0\equiv \int\limits_{0}^{1} \mu
\Big[h(0,\mu)-h(0,-\mu)\Big]d\mu=0.
$$

Taking into account the relation (2.16), we obtain that
$
A_0=-2A_1/3.
$
With the help of this relation we can rewrite the condition
(2.16) in the following form:
$$
h(0,\mu)=h(0,-\mu)+A_1(\mu-\dfrac{2}{3}), \qquad 0<\mu<1.
\eqno{(3.1)}
$$

We consider the momentum flow of the electrons which are moving to
the boundary. According to (3.1) we have:
$$
P_i=P_r-\dfrac{1}{36}A_1.
\eqno{(3.2)}
$$

It is easy to see further that
$$
 P_s={A_s}/{3}.
\eqno{(3.3)}
$$

With the help of the formula (3.2) we will rewrite the definition of
the accommodation coefficient (2.12) in the form:
$$
\alpha_pP_r-\alpha_p\dfrac{A_s}{3}+\dfrac{A_1}{36}(1-\alpha_p)=0.
\eqno{(3.4)}
$$

Let us consider the condition (2.15). From this condition we obtain
that
$$
A_s=-2\int\limits_{-1}^{0}\mu h(0,\mu)d\mu=2\int\limits_{0}^{1}
\mu h(0,-\mu)d\mu.
$$

Using the condition (3.1), we then get
$$
A_s=2\int\limits_{0}^{1}\mu h(0,\mu)d\mu. \eqno{(3.5)}
$$

Now with the help of the second equality from (2.13) and (3.4) we
rewrite the relation (3.3) in the integral form:
$$
\alpha_p\int\limits_{0}^{1}\Big(\mu^2-\dfrac{2}{3}\Big)h(0,\mu)d\mu=
-\dfrac{1}{36}(1-\alpha_p)A_1. \eqno{(3.6)}
$$
Now the boundary problem consists of the equations (2.8) and (2.9)
and boundary conditions (2.10), (3.1) and (3.6).\\

\begin{center}\bf
4. SEPARATION OF VARIABLES AND CHARACTERISTICAL SYSTEM
\end{center}

Application of the general Fourier method of the separation of
variables in several steps results in the following substitution:
$$
h_\eta(x,\mu)=\exp(-\dfrac{z_0x}{\eta})\Phi(\eta,\mu),\qquad
e_\eta(x)=\exp(-\dfrac{z_0x}{\eta})E(\eta),
\eqno{(4.1)}
$$
where $\eta$ is the spectrum parameter or the parameter of
separation, which is complex in general.

We substitute the equalities (4.1) into the equations (2.8) and
(2.9). We obtain the following characteristic system of equations:
$$
z_0(\eta-\mu)\Phi(\eta,\mu)=\eta\mu E(\eta)+\dfrac{\eta}{2}
\int\limits_{-1}^{1}\Phi(\eta,\mu')d\mu',
\eqno{(4.2)}
$$
$$
z_0E(\eta)=-\dfrac{3}{\varepsilon^2}\cdot\dfrac{\eta}{2}\int\limits_{-1}^{1}
\Phi(\eta,\mu')d\mu'.
\eqno{(4.3)}
$$

Let us introduce the designations:
$$
\gamma=\dfrac{\omega}{\omega_p}-1, \quad
\eta_1^2=\dfrac{\varepsilon^2z_0}{3},\quad
z_0=1-i\dfrac{1+\gamma}{\varepsilon}, \quad
c=2\eta_1^2z_0.
$$

Substituting the integral from the equation (4.3) into (4.2), we
come to the following system of equations:
$$
(\eta-\mu)\Phi(\eta,\mu)=\dfrac{E(\eta)}{z_0}(\eta\mu-\eta_1^2),\quad
-\eta_1^2E(\eta)=\dfrac{\eta}{2}\int\limits_{-1}^{1}
\Phi(\eta,\mu')d\mu'.
\eqno{(4.4)}
$$

Solution of the system (4.4) depends essentially on the condition if
the spectrum parameter $\eta$ belongs to the interval $-1<\eta<1$.
In connection with this the interval $-1<\eta<1$ we will call as
continuous spectrum of the characteristic system.

Let the parameter $\eta\in (-1,1)$. Then from the equations (4.4) in
the class of general functions we will find eigenfunction
corresponding to the continuous spectrum:
$$
\Phi(\eta,\mu)=F(\eta,\mu)\dfrac{E(\eta)}{z_0},
\eqno{(4.5)}
$$
where
$$
F(\eta,\mu)=P\dfrac{\mu\eta-\eta_1^2}{\eta-\mu}-c
\dfrac{\lambda(\eta)}{\eta}\delta(\eta-\mu).
\eqno{(4.6)}
$$

In the equation (4.6) $\delta(x)$ is the delta--function of Dirac,
the symbol $Px^{-1}$ means the principal value of the integral under
integrating of the expression $x^{-1}$, the fuction $\lambda(z)$ is
called as dispersion function of the problem,
$$
\lambda(z)=1+\dfrac{z}{c}\int\limits_{-1}^{1}
\dfrac{\eta_1^2-z\mu}{\mu-z}d\mu.
\eqno{(4.7)}
$$

The function (4.5) is called eigenfunction of the continuous
spectrum, since the spectrum parameter $\eta$ fills out the
continuum $(-1,+1)$ compactly. The eigensolutions of the given
problem can be found from the equalities (4.7). The dispersion
function $\lambda(z)$ we express in the terms of
the Case \cite{Case} dispersion function:
$$
\lambda(z)=1-\dfrac{1}{z_0}+\dfrac{1}{z_0}\Big(1-
\dfrac{z^2}{\eta_1^2}\Big)\lambda_c(z),
$$
where
$$
\lambda_c(z)=1+\dfrac{z}{2}\int\limits_{-1}^{1}\dfrac{d\tau}{\tau-z}=
\dfrac{1}{2}\int\limits_{-1}^{1}\dfrac{\tau\,d\tau}{\tau-z}
$$
is the Case dispersion function  \cite{Case}.

The boundary values of the dispersion function from above and below
the contour are calculated according to the Sokhotsky formulas
$$
\lambda^{\pm}(\mu)=\lambda(\mu)\pm \dfrac{i \pi\mu}
{2\eta_1^2z_0}(\eta_1^2-\mu^2),\quad -1<\mu<1,
$$
from where we have
$$
\lambda^+(\mu)-\lambda^-(\mu)=\dfrac{i \pi}{\eta_1^2z_0}
\,\mu(\eta_1^2-\mu^2),
$$
$$
\dfrac{\lambda^+(\mu)+\lambda^-(\mu)}{2}=\lambda(\mu),\quad-1<\mu<1,
$$
where
$$
\lambda(\mu)=1+\dfrac{\mu}{2\eta_1^2z_0} \int\limits_{-1}^{1}
\dfrac{\eta_1^2-\eta^2}{\eta-\mu}\,d\eta,
$$
and the integral in this equality is understood as singular in terms
of the principal value by Cauchy. Besides that, the function
$\lambda(\mu)$ can be represented in the following form:
$$
\lambda(\mu)=1-\dfrac{1}{z_0}+
\dfrac{1}{z_0}\Big(1-\dfrac{\mu^2}{\eta_1^2}\Big)\lambda_c(\mu),
\quad
\lambda_c(\mu)=1+\dfrac{\mu}{2}\ln\dfrac{1-\mu}{1+\mu}.
$$\\

\begin{center}\bf
5. EIGENFUNCTIONS OF THE DISCRETE SPECTRUM AND PLASMA WAVES
\end{center}

According to the definition, the discrete spectrum of the
characteristic equation is a set of zeroes of the dispersion
equation
$$ \lambda(z)/{z}=0.
\eqno{(5.1)}
$$

We start to search zeroes of this equation. Let us expand take Laurent
series of the dispersion function:
$$
\lambda(z)=\lambda_\infty+\dfrac{\lambda_2}{z^2}+
\dfrac{\lambda_4}{z^4}+\cdots,\qquad |z|>1.
\eqno{(5.1)}
$$

Here
$$
\lambda_\infty \equiv\lambda(\infty)=
1-\dfrac{1}{z_0}+\dfrac{1}{3z_0\eta_1^2}=
\dfrac{2\gamma+i\varepsilon+\gamma(\gamma+i\varepsilon)}{(1+\gamma+
i\varepsilon)^2},
$$
$$
\lambda_2=-\dfrac{1}{z_0}\Big(\dfrac{1}{3}-\dfrac{1}{5\eta_1^2}\Big)=
-\dfrac{9+5i\varepsilon(1+\gamma+i\varepsilon)}{15(1+\gamma+
i\varepsilon)^2},
$$
$$
\lambda_4=-\dfrac{1}{z_0}\Big(\dfrac{1}{5}-\dfrac{1}{7\eta_1^2}\Big)=
-\dfrac{15+7i\varepsilon(1+\gamma+i\varepsilon)}{35(1+\gamma+
i\varepsilon)^2}.
$$

One can easily see that in collisional plasma (i.e. when
$\varepsilon>0$) the coefficient $\lambda_\infty\ne 0$.
Consequently, the dispersion equation has infinity as a zero
$\eta_i=\infty$, to which the discrete eigensolutions of the given
system correspond: $h_\infty(x,\mu)=\mu/z_0,\;e_\infty(x)=1$.

This solution is naturally called as mode of Drude. It describes the
volume conductivity of metal, considered by Drude
(see, for example, \cite{10}).

Let us consider the question of the plasma mode existance in
details. We find finite complex zeroes of the dispersion function.
We use the principle of argument. We take the contour
\;$\Gamma_\varepsilon^+=\Gamma_R\cup \gamma_\varepsilon$, (see Fig.
1), which is passed in the positive direction and which bounds the
biconnected domain $D_R$. This contour consists of the circumference
$\{\Gamma_R: |z|=R,\;R=1/\varepsilon, \varepsilon>0\}$, and the
contour $\gamma_\varepsilon$, which includes the cut $[-1,+1]$, and
stands at the distance of $\varepsilon$ from it.

According to the principle of argument the number \cite{11} of
zeroes $N$ in the area $D_\varepsilon$ equals to:
$$
N=\dfrac{1}{2\pi
i}\oint\limits_{\Gamma_\varepsilon}d\,\ln\lambda(z).
$$

\begin{figure}[t]
\begin{center}
\includegraphics[width=11cm, height=15cm]{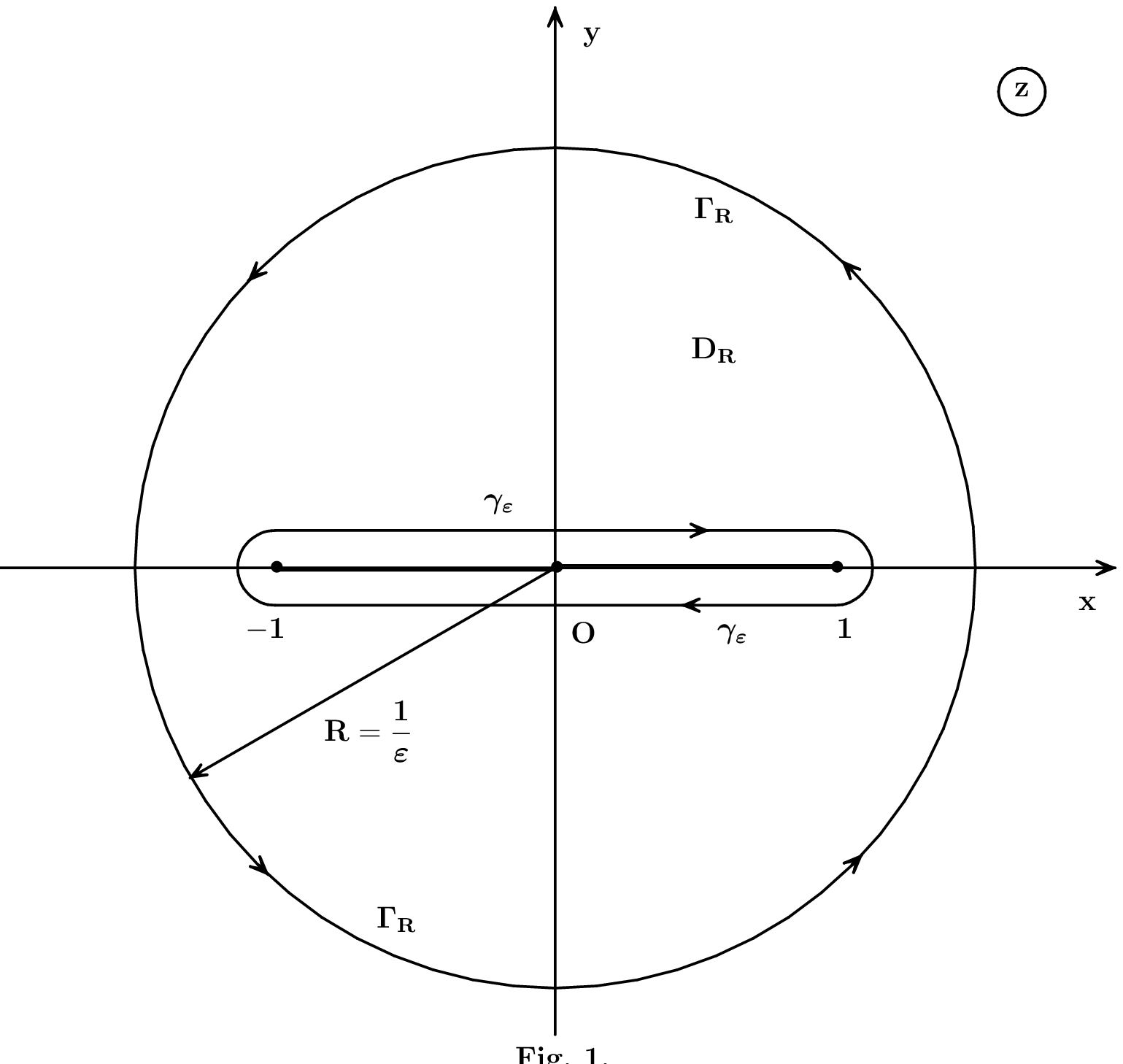}
\end{center}
\end{figure}
Considering the limit in this equality when $\varepsilon\to 0$ and
taking into account that the dispersion function is analytic in the
neighbourhood of the infinity, we obtain that
$$
N=\dfrac{1}{2\pi i}\int\limits_{-1}^{1}d\,\ln
\dfrac{\lambda^+(\tau)}{\lambda^-(\tau)}=
\dfrac{1}{\pi i}\int\limits_{0}^{1}d\,\ln
\dfrac{\lambda^+(\tau)}{\lambda^-(\tau)}=\dfrac{1}{\pi}\arg
\dfrac{\lambda^+(\tau)}{\lambda^-(\tau)}\Bigg|_0^1.
\eqno{(5.2)}
$$

Consider the curve
$$
\gamma=\big\{z:\quad\;z=G(\tau),\;0\leqslant \tau\leqslant+1\big\},
$$
where $ G(\tau)=\lambda^+(\tau)/\lambda^-(\tau). $
It is obvious that $G(0)=1,\; \lim\limits_{\tau\to 1}G(\tau)=1$.
Hence, according to (5.3), the number of zeroes $N$ equals to the
double number of turns of the curve $\gamma$ round the point of
origin, i.e. $ N=2\varkappa(G)$,\; $\varkappa(G)={\rm
Ind}_{[0,+1]}G(\tau). $

Thus, the number of zeroes of the dispersion function belonging to
the complex plane out of the segment $[-1,1]$ of the real axis
equals to the double index of the function $G(\tau)$, calculated on
the semi-axis $[0,+1]$.

We take on the plane $(\gamma,\varepsilon)$ (see Fig. 2) the curve
$L$, determined by equations:
$\gamma=-1+\sqrt{L_1(\tau)},\;\varepsilon=\sqrt{L_2(\tau)},\;
0\leqslant\tau\leqslant 1,$  where
$$
L_1(\tau)=-\dfrac{3\mu^2[s^2+\lambda_0(1+\lambda_0)]^2}
{\lambda_0[s^2+(1+\lambda_0)^2]},\quad
L_2(\tau)=-\dfrac{3\tau^2s^2}{\lambda_0
[s^2+(1+\lambda_0)^2]}.
$$
Here
$$
\lambda_0(\tau)=1+\dfrac{\tau}{2}\ln\dfrac{1-\tau}{1+\tau},
\quad s(\tau)=\dfrac{\pi}{2}\tau.
$$

In the same way, as was shown in the work \cite{12}, we can prove
that if $(\gamma,\varepsilon)\in D^+$, then the index of the problem
equals to unity, i.e. $N=2$ is the number of zeroes which equals to
two, and if $(\gamma,\varepsilon)\in D^-$, then the index of the
problem equals to zero, i.e. $N=0$.

We would like to notice, that in the work \cite{13} (see also
\cite{12}) the method of examination of the boundary mode in the
case when $(\gamma,\varepsilon)\in L$ was developed.

\begin{figure}
\begin{center}
\includegraphics{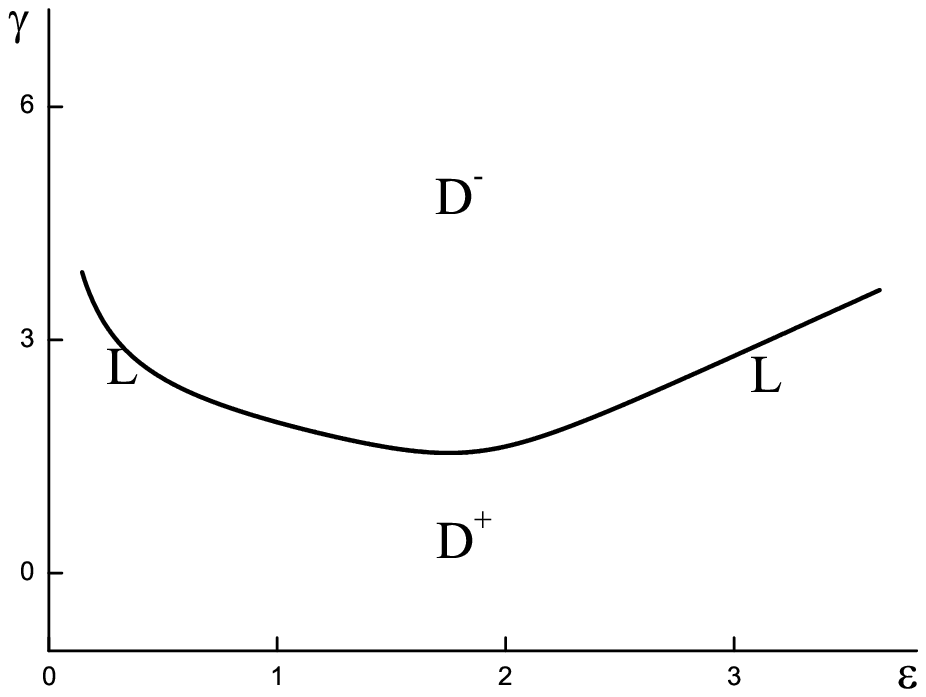}
\end{center}
\begin{center}
{\bf Fig. 2.}
\end{center}
\end{figure}

Since the dispersion function is even its zeroes differ from each
other by sign. We designate these zeroes as following $\pm \eta_0$,
by $\eta_0$ we take the zero which satisfies the condition $\Re
\eta_0>0$. The following solution corresponds to the zero $\eta_0$:
$$
h_{\eta_0}(x,\mu)=\exp(-\dfrac{z_0x}{\eta_0})
\dfrac{E_2}{z_0}\dfrac{\eta_0\mu-\eta_1^2}{\eta_0-\mu},\qquad
e_{\eta_0}(x)=\exp(-\dfrac{z_0x}{\eta_0})E_2.
\eqno{(5.2)}
$$

This solution is naturally called as mode of Debay (this is plasma
mode). In the case of low frequencies it describes well-known
screening of Debay \cite{3}. The external field penetrates into
plasma on the depth of $r_D,\; r_D$ is the raduis of Debay. When the
external field frequancies are close to Langmuir frequencies, the
mode of Debay describes plasma oscillations (see, for instance,
\cite{3,10}).

From the equalities  (2.2) for the wave $e_-(x,t)$ with the help of
(2.6) and the equality (4.1) follows the relation between the wave
number $k$ and the zero of the dispersion function
$\eta_0(\omega,\nu)$: $
i\dfrac{kx}{\varepsilon}=-\dfrac{z_0x}{\eta_0}, $ hence $ \eta_0
\equiv \eta_0(\gamma,\varepsilon)=\dfrac{1+\gamma}{k}+i
\dfrac{\varepsilon}{k}. $

The equalities (2.5) and (2.6) jointly with (5.2) mean that the
reflected wave corresponds to the zero $\eta_0$:
$$
H_{\eta_0}=\dfrac{\eta_1^2-\eta_0\mu}
{z_0(\mu-\eta_0)}\exp\Big[i\Big(\dfrac{kx}{\varepsilon}-\omega
t\Big)\Big],\quad
e_{\eta_0}=\exp\Big[i\Big(\dfrac{kx}{\varepsilon}-\omega
t\Big)\Big],
$$
and the wave incoming to the plasma boundary corresponds to the
symmetric zero $-\eta_0$:
$$
H_{-\eta_0}=\dfrac{\eta_1^2+\eta_0\mu}
{z_0(\mu+\eta_0)}\exp\Big[-i\Big(\dfrac{kx}{\varepsilon}+\omega
t\Big)\Big],\quad
e_{-\eta_0}=\exp\Big[-i\Big(\dfrac{kx}{\varepsilon}+\omega
t\Big)\Big],
$$\\

\begin{center}\bf
6. EXPANSION IN THE TERMS OF EIGENFUNCTIONS
\end{center}

In the work \cite{8} it was shown that from the non--flowing
condition (2.11) and the condition on the electric field (2.10) it
results that the trivial (equal to zero) solution of the present
problem corresponds to the point $\eta_i=\infty$.

We will show that the system of equations (2.8) and (2.9) with the
boundary conditions (3.1), (3.5) and (2.10) has the solution which
can be represented as an expansion by the eigenfunctions of the
characteristic system:
$$
h(x,\mu)=\dfrac{E_2}{z_0}
 \dfrac{\eta_0\mu-\eta_1^2}{\eta_0-\mu}
 \exp(i\dfrac{kx}{\varepsilon})+
 \dfrac{E_1}{z_0}
 \dfrac{\eta_0\mu+\eta_1^2}{\eta_0+\mu}
 \exp(-i\dfrac{kx}{\varepsilon})+
 $$
 $$
 +\dfrac{1}{z_0}
 \intl_{0}^{1}\exp\Big(-z_0\dfrac{x}{\eta}\Big)
 F(\eta,\mu)E(\eta)\,d\eta,
 \eqno{(6.1)}
 $$
$$
 e(x)=E_2\exp(i\dfrac{kx}{\varepsilon})+
 E_1\exp(-i\dfrac{kx}{\varepsilon})
 +\intl_{0}^{1}
 \exp\Big(-z_0\dfrac{x}{\eta}\Big)E(\eta)\,d\eta.
 \eqno{(6.2)}
 $$

 Here $E_1$ is given, and $E_2$ is unknown coefficient.
 Both of the variables (amplitudes of Debay) correspond to
 the discrete spectrum, $E(\eta)$ is unknown function,
 which is called eigenfunction of continuous spectrum.

Our purpose is to find the coefficient of the continuous spectrum
and the relation which connects the coefficients of the discrete
spectrum.

Let us substitute the expansion (6.1) into the boundary condition
(3.1). We get the following equation in the interval $0<\mu<1$:
 $$
\intl_{0}^{1}\Big[F(\eta,\mu)-
 F(\eta,-\mu)\Big]E(\eta)d\eta+ (E_1+E_2)\varphi_0(\mu)=
-\dfrac{2}{3}z_0A_1+z_0A_1\mu.
 \eqno{(6.3)}
 $$

 Here
 $$
 \varphi_0(\mu)=\dfrac{\eta_1^2-\eta_0\mu}{\mu-\eta_0}+
 \dfrac{\eta_1^2+\eta_0\mu}{\mu+\eta_0}.
 $$

 Extending the function $E(\eta)$ into the interval $(-1,0)$
 evenly we transform the equation (6.3) to the following form:
 $$
\intl_{-1}^{1}
 F(\eta,\mu)E(\eta)d\eta+(E_1+E_2)\varphi_0(\mu)-
z_0A_1\mu=-\dfrac{2}{3}z_0A_1\sign \mu.
 \eqno{(6.4)}
 $$

Let us substitute the eigenfunctions of the continuous spectrum into
the equation (6.4). We obtain singular integral equation with Cauchy
kernel in the interval $(-1,1)$:
 $$
 (E_1+E_2)\varphi_0(\mu)+\intl_{-1}^{1}
 \dfrac{\eta\mu-\eta_1^2}{\eta-\mu}E(\eta)d\eta-
c\dfrac{\lambda(\mu)}{\mu}E(\mu)-
 z_0A_1\mu=$$$$=-\dfrac{2}{3}z_0A_1\sign \mu.
\eqno{(6.5)}
 $$\\

\begin{center}\bf
7. SOLUTION OF THE SINGULAR EQUATION
\end{center}

 We introduce the auxiliary function
 $$
 M(z)=\intl_{-1}^{1}\dfrac{\eta z-\eta_1^2}{\eta-z}
 E(\eta)d\eta,
\eqno{(7.1)}
 $$
 the boundary values of which on the real axis above and below it
 are related by the Sokhotsky formulas:
 $$
 M^+(\mu)-M^-(\mu)=2\pi i (\mu^2-\eta_1^2)E(\mu).
\eqno{(7.2)}
 $$
$$
\dfrac{M^+(\mu)+M^-(\mu)}{2}=M(\mu),\qquad -1<\mu<+1, \eqno{(7.3)}
$$
where
$$
M(\mu)=\intl_{-1}^{1}\dfrac{\eta \mu-\eta_1^2}{\eta-\mu}
 E(\eta)d\eta,
$$
and the singular integral in this equality is understood as singular
in the sense of principal value of Cauchy.

With the help of the Sokhotsky formulas for the dispersion and
auxiliary function we reduce the equation (6.5) to the boundary
condition of the problem of determination of analytic function by
its jump on the contour:
 $$
 \lambda^+(\mu)[M^+(\mu)+(E_1+E_2)\varphi_0(\mu)-z_0A_1\mu]-
\lambda^-(\mu)[M^-(\mu)+(E_1+E_2)\varphi_0(\mu)-$$$$-z_0A_1\mu]=
\dfrac{i\pi }{3\eta_1^2}A_1\mu(\eta_1^2-\mu)\sign \mu,
\quad  -1<\mu<1.
 $$

This equation has general solution (see \cite{11}):
$$
\lambda(z)[\varphi(z)(E_1+E_2)+M(z)-z_0A_1z]
=\dfrac{2z_0A_1}{3c}\int\limits_{-1}^{1}
\dfrac{\mu(\mu^2-\eta_1^2)\sign \mu}{\mu-z}d\mu+C_1z,
$$
where $C_1$ is an arbitrary constant.

Let us introduce auxiliary function
$$
T(z)=\dfrac{1}{c}\int\limits_{-1}^{1}
\dfrac{\mu(\mu^2-\eta_1^2)\sign \mu}{\mu-z}d\mu.
$$

Then from the general solution we can easy find $M(z)$:
$$
M(z)=-(E_1+E_2)\varphi(z)+z_0A_1z+
\dfrac{2}{3}z_0A_1\dfrac{T(z)}{\lambda(z)}+\dfrac{C_1z}{\lambda(z)}.
\eqno{(7.4)}
$$

Let us eliminate the pole of the solution (7.4) in the infinity. We
get that
$$
C_1=-z_0A_1\lambda_\infty.
$$

Poles in the points $z=\pm \eta_0$ can be eliminated with the help
of one equality since the functions constituting the general
solution are uneven:
$$
z_0A_1=\dfrac{(E_1+E_2)\lambda'(\eta_0)(\eta_1^2-\eta_0^2)}
{(2/3)T(\eta_0)-\lambda_\infty \eta_0}. \eqno{(7.5)}
$$

We substitute the expansion (6.1) for the function $h(x,\mu)$ to the
integral boundary condition (3.5). We get the following equation:
$$
E_1 m(-\eta_0)+E_2m(\eta_0)+\int\limits_{0}^{1}m(\eta)E(\eta)d\eta
=-\dfrac{1}{36}z_0A_1\dfrac{1-\alpha_p}{\alpha_p}. \eqno{(7.6)}
$$

In (7.6) the following designations were introduced:
$$
m(\pm \eta_0)=\int\limits_{0}^{1}\Big(\mu^2-\dfrac{2}{3}\Big)
F(\pm \eta_0,\mu)d\mu,\quad
m(\eta)=\int\limits_{0}^{1}\Big(\mu^2-\dfrac{2}{3}\Big)
F(\eta,\mu)d\mu.
$$

The coefficient of the continuous spectrum we will find from the
Sokhotsky formula (7.1) after the substitution of the general
solution (7.4) into it:
$$
E(\eta)=\dfrac{1}{2\pi i (\eta^2-\eta_1^2)}
\Bigg[\dfrac{2}{3}\Big[\dfrac{T^+(\eta)}{\lambda^+(\eta)}-
\dfrac{T^-(\eta)}{\lambda^-(\eta)}\Big]-\lambda_\infty \eta
\Big[\dfrac{1}{\lambda^+(\eta)}-
\dfrac{1}{\lambda^-(\eta)}\Big]\Bigg].
\eqno{(7.7)}
$$

Let us notice that under the transition through the positive part of
the cut $(0,1)$ functions $T(z)$ and $\lambda(z)$ make jumps, which
differ only by sign. Indeed, let us represent the formula for $T(z)$
in the following form:
$$
T(z)=\dfrac{z}{c}\int\limits_{0}^{1}
(\mu^2-\eta_1^2)\Bigg[\dfrac{1}{\mu-z}+\dfrac{1}{\mu+z}\Bigg]d\mu.
$$

This integral can be calculated easily in explicit form. On the cut
this integral is calculated according to the following formula :
$$
T(\eta)=\dfrac{\eta}{c}\Big[1+(\eta^2-\eta_1^2)
\ln\Big(\dfrac{1}{\eta^2}-1\Big)\Big], \qquad -1<\eta<+1.
$$

Now from the Sokhotsky formula for the difference of boundary values
we obtain that under condition $0<\eta<1$ the following equality
takes place:
$$
\lambda^+(\eta)-\lambda^-(\eta)=\lambda(\eta)\pm
\dfrac{i\pi \eta(\eta_1^2-\eta^2)}{c},
$$
$$
T^+(\eta)-T^-(\eta)=T(\eta)\pm
\dfrac{i\pi \eta (\eta^2-\eta_1^2)}{c}.
$$

Now one can naturally find that
$$
T^+(\eta)\lambda^-(\eta)-T^-(\eta)\lambda^+(\eta)=
2(T(\eta)+\lambda(\eta))\cdot
\dfrac{i\pi \eta (\eta^2-\eta_1^2)}{c},
$$
$$
\lambda^-(\eta)-\lambda^+(\eta)=
2 \dfrac{i\pi \eta (\eta^2-\eta_1^2)}{c}.
$$

With the help of the last relations we find the coefficient of the
continuous spectrum from (7.5):
$$
E(\eta)=z_0A_1 Q(\eta), \quad \text{where}\quad Q(\eta)=
\dfrac{\frac{2}{3}\eta[T(\eta)+\lambda(\eta)]-\lambda_\infty\eta^2}
{c\lambda^+(\eta)\lambda^-(\eta)}. \eqno{(7.8)}
$$

We introduce the integral
$$
T_0(z)= \dfrac{1}{c}\int\limits_{0}^{1}
\dfrac{\eta^2-\eta_1^2}{\eta-z}d\eta.
$$

It is evident that in the complex plane this integral is calculated
by the following formula:
$$
T_0(z)=\dfrac{z}{c}\Big[\dfrac{1}{2}+z+(z^2-\eta_1^2)
\ln\Big(\dfrac{1}{z}-1\Big)\Big].
$$

With the help of this function we represent the dispersion function
in the form: $ \lambda(z)=1-zT_0(z)+zT_0(-z), $ the function $T(z)$
we also express in terms of this integral: $ T(z)=zT_0(z)+zT_0(-z).
$ The sum of two last expressions equals to: $
\lambda(z)+T(z)=1+2zT_0(-z). $ Let us note that the integral $T(-z)$
is not singular on the cut $0<\eta<1$. The sum
$\lambda(\eta)+T(\eta)$ on the cut $0<\eta<1$ is calculated in
explicit form without applying to integrals:
$$
\lambda(\eta)+T(\eta)=1+\dfrac{1}{c}\Big[\eta-
2\eta^2+2\eta(\eta^2-\eta_1^2)\ln(1/\eta+1)\Big].
$$

We calculate the integrals $m(\pm \eta_0)$ and $m(\eta)$ in explicit
form. The integrals  $m(\pm \eta_0)$  can be determined easily:
$$
m(\pm\eta_0)=-(\eta_0^2-\eta_1^2)\Big[-\dfrac{1}{6}+\eta_0+
\eta_0(\eta_0-\dfrac{2}{3})\ln(\dfrac{1}{\eta_0}-1)\Big].
$$

Let us find the integral $m(\eta)$. We have:
$$
m(\eta)=
\int\limits_{0}^{1}\Big(\mu^2-\dfrac{2}{3}\mu\Big)(\eta_1^2-\eta \mu)
\dfrac{d\mu}{\mu-\eta}-2\eta_1^2z_0(\eta-\dfrac{2}{3})
\lambda(\eta)\theta_+(\eta).
$$

Here $\theta_+(\eta)$ is the characteristic function of the interval
 $0<\eta<1$, i.e.
$$
\theta_+(\eta)= \Big\{
\begin{array}{l}
  1,\quad 0<\eta<1,  \\
  0,\quad -1<\eta<0.
\end{array}
$$

Computating the integral in the preceding equality we obtain that
the integral $m(\eta)$ is calculated by the formula:
$$
m(\eta)=(\dfrac{1}{6}-\eta)(\eta^2-\eta_1^2)+
(\eta-\dfrac{2}{3})\Big[-c+2\eta^2-
\eta(\eta^2-\eta_1^2)f_+(\eta)\Big],
$$
where
$$
f_+(\eta)=\left\{
\begin{array}{l}
  \ln\dfrac{1+\eta}{\eta},\quad 0<\eta<1,  \\
  \ln\dfrac{1-\eta}{\eta},\quad -1<\eta<0.
\end{array}\right.
$$

Now the equation (7.6) with the help (7.8) we rewrite in the form:
$$
E_1m(-\eta_0)+E_2m(\eta_0)=-z_0A_1\Big(\dfrac{1-\alpha_p}{36\alpha_p}+
Q_m\Big), \eqno{(7.9)}
$$
where
$$
Q_m=\int\limits_{0}^{1}m(\eta)Q(\eta)d\eta.
$$

After that substituting the variable $z_0A_1$ into the equation
(7.9) according to (7.5), we derive the equation:
$$
E_1m(-\eta_0)+E_2m(\eta_0)=-\Big(\dfrac{1-\alpha_p}{36\alpha_p}+
Q_m\Big)\dfrac{E_1+E_2}{\frac{2}{3}T(\eta_0)-\lambda_\infty\eta_0},
$$
from which we can find the amplitude required $E_2$:
$$
E_2=-\dfrac{\alpha_pm(-\eta_0)A(\eta_0)+B(\eta_0)C(\alpha_p)}
{\alpha_pm(\eta_0)A(\eta_0)+B(\eta_0)C(\alpha_p)}E_1. \eqno{(7.10)}
$$

We introduced the following designations in the formula (7.10):
$$
A(\eta_0)=\dfrac{2}{3}T(\eta_0)-\lambda_\infty \eta_0,\quad
C(\alpha_p)=\dfrac{1-\alpha_p}{36}+\alpha_pQ_m,
$$
$$
B(\eta_0)=(\eta_1^2-\eta_0^2)\lambda'(\eta_0).
$$

Thus, all the coefficients of the expansions (6.1) and (6.2) are
determined unambiguously, and this completes the proof of these
expansions.

It is seen from the equality (7.10) that under condition
$\alpha_p=0$ we have: $E_2=-E_1$, i.e. under condition of pure
specular reflection of electrons from the boundary the wave
reflectance equals to unity: $R=1$, and the phase shift of the
incoming and reflected waves is equal  to $180^\circ$, i.e.
$\phi=\pi,$ from where we see that $\arg E_2=\arg E_1+\pi.$

Let us represent the formula (7.10) in the form which is more
convenient for numerical analysis. Let us designate the ratio of
amplitudes as $K,\;K=E_2/E_1$, the
$$
K
=-1+\dfrac{\alpha_p[m(\eta_0)-m(-\eta_0)]}
{\alpha_pm(\eta_0)+C(\alpha_p)D(\eta_0)},
\eqno{(7.9)}
$$
where
$$
D(\eta_0)=\dfrac{B(\eta_0)}{A(\eta_0)}=
\dfrac{(\eta_1^2-\eta_0^2)\lambda'(\eta_0)}{(2/3)T(\eta_0)-
\lambda_\infty\eta_0}.
$$\\

\begin{center}\bf
  8. LONG WAVE LIMIT
\end{center}

For study of the incoming wave reflectance  $R=|K|^2$ and the phase
shift $\phi=\arg K$ we will use the formula (7.11).

We consider the dispersion equation with small values of the wave
number $k$:
$$
\lambda\Big(i\dfrac{\varepsilon z_0}{k}\Big)=\lambda_\infty-
\dfrac{\lambda_2k^2}{\varepsilon^2z_0^2}=0.
\eqno{(8.1)}
$$

We assume the frequency  $\omega$ is complex:
$\omega=\omega_0+i\omega_1$. Then the quantity $\gamma$ is complex
also: $\gamma=\gamma_0+i\gamma_1$. Here
$\gamma_0=\omega_0/\omega_p-1,\; \gamma_1=\omega_1/\omega_p$. From
the equation (8.1) we find that when $k$ is small
$\gamma_0=0.3k^2,\;\gamma_1=-0.5\varepsilon$. We express the
parameters of the problem in terms of $k$ and $\varepsilon$:
$\lambda_\infty=0.6k^2(1-i\varepsilon)$,
$$
z_0=\dfrac{1}{2}-\dfrac{1+0.3k^2}{\varepsilon},\quad
\eta_1^2=-i\dfrac{\varepsilon}{3}(1+0.3k^2), \quad
\eta_0=\dfrac{1+0.3k^2+i0.5\varepsilon}{k}.
$$

With the help of these parameters let us carry out the study of the
reflectance and the phase shift in long wave limit (when $k$ is
small by magni\-tu\-de).

On the Fig. 3 one can see the dependence of the reflectance $R$ on
the wave number $k$ for the case when $\varepsilon=10^{-2}$. The
curves $1,2,3$ correspond to the following values of the
accommodation coefficient $\alpha_p=0.1,0.5,1.0$. The curve $4$
corresponds to the diffuse boundary conditions (see \cite{7}). From
the graph it is seen that when $k$ takes on small values the curve
$3$ (corresponding to $\alpha_p=1$) coincides practically with the
curve $4$ (to which corresponds the case $q=1$), which was obtained
by the linearization of the reflectance value by $k$. Taking into
account that under $\alpha_p=0$ the reflectance is equal to
reflectance for specular boundary conditions (i.e. when $q=0$) we
can conclude that specular accommodative boundary conditions
approximate specular and diffuse boundary conditions under
$\alpha_p=q$ very well.

On the Fig. 4 the dependence of the reflectance $R$ on the quantity
$\varepsilon$ for the case $\alpha_p=1$ is represented. The curves
 $1,2,3,4,5$ correspond to the following values of the wave number
$k=0.001,0.01,0.05,0.1,0.2$. The more accurate analysis shows that
with the growth of the accommodation coefficient the value of the
reflectance decreases.

\begin{figure}[t]
\begin{center}
\includegraphics[width=0.49\textwidth, height=7cm]{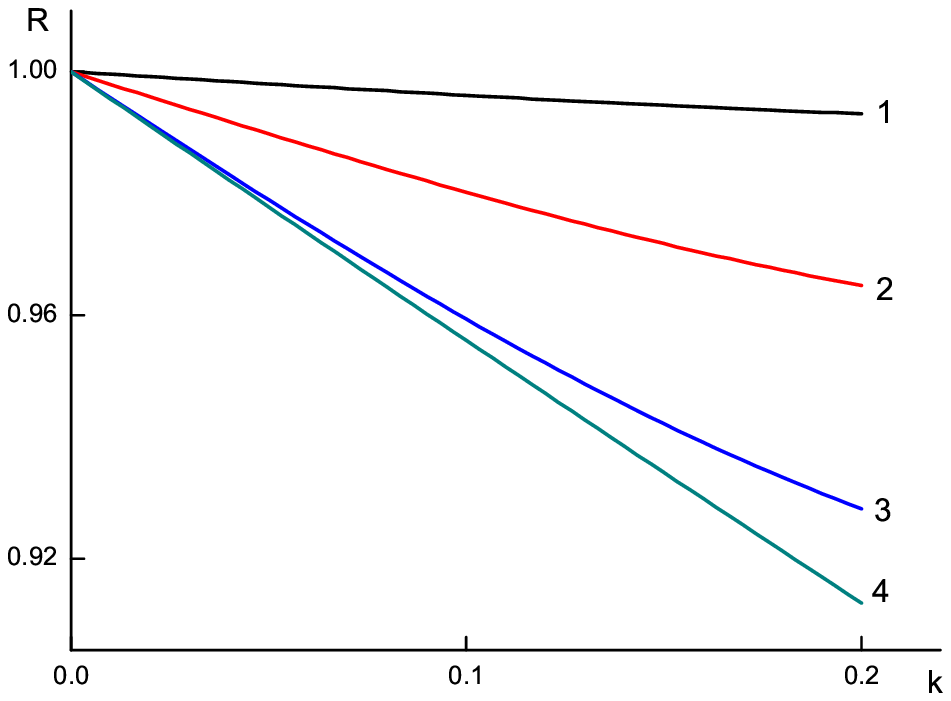}
\hfill
\includegraphics[width=0.49\textwidth, height=7cm]{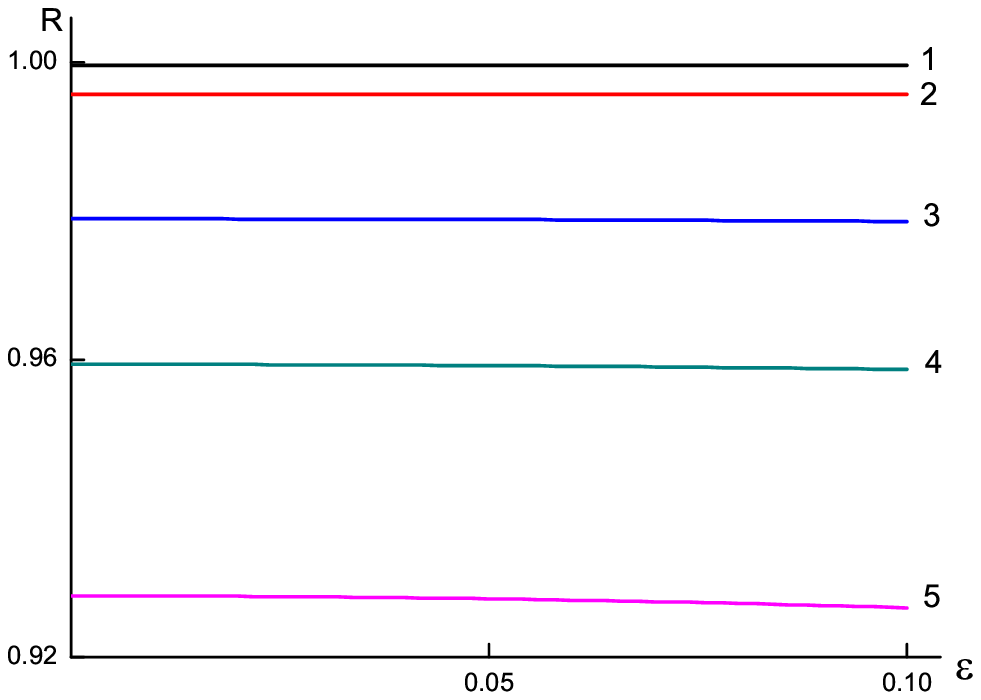}
\\
\parbox[t]{0.47\textwidth}{\hspace{3cm}{Fig. 3.}}
\hfill
\parbox[t]{0.47\textwidth}{\hspace{3cm}{Fig. 4.}}
\end{center}
\end{figure}~
\begin{figure}[h]
\begin{center}
\includegraphics[width=0.47\textwidth, height=7cm]{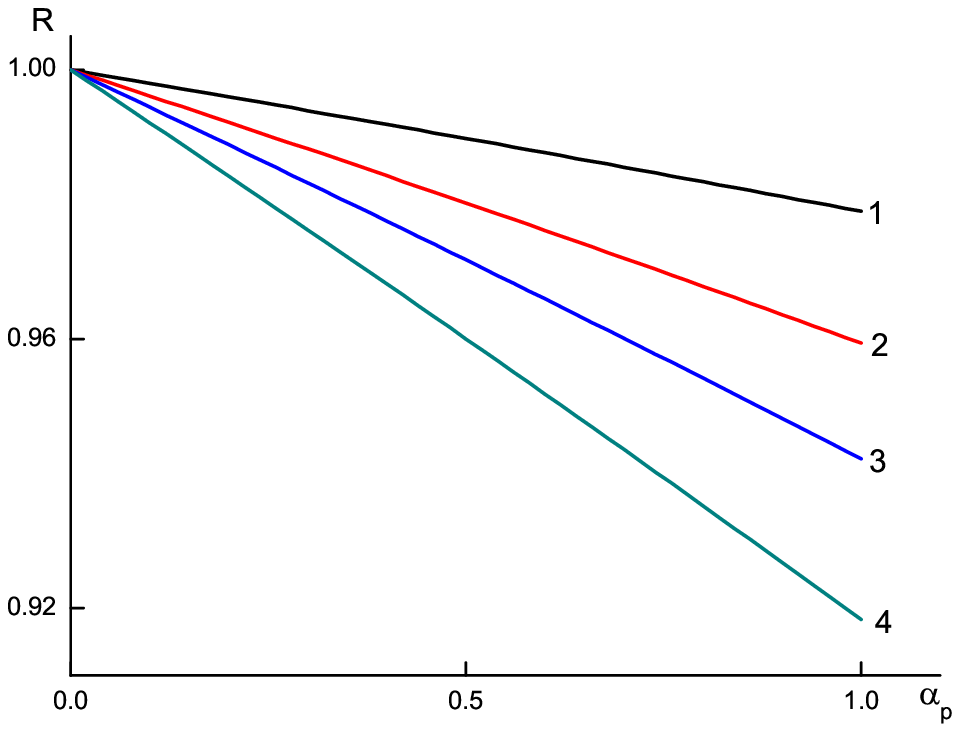}
\hfill
\includegraphics[width=0.47\textwidth, height=7cm]{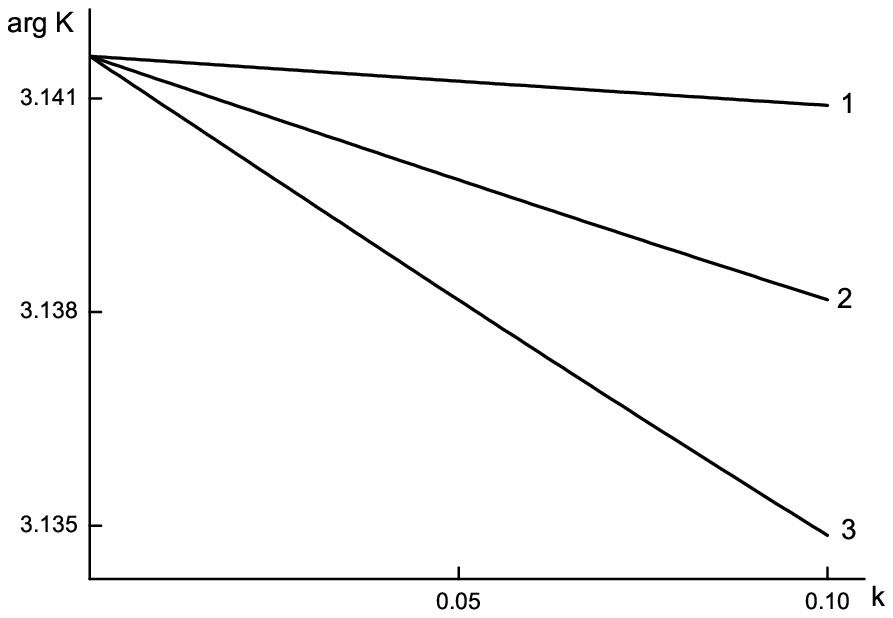}
\\
\parbox[t]{0.47\textwidth}{
\hspace{3cm}{Fig. 5. }}
\hfill
\parbox[t]{0.47\textwidth}{
\hspace{3cm}{Fig. 6. }}
\end{center}
\end{figure}

On the Fig. 5 the dependence of the reflectance $R$ on the value of
the accommodation coefficient $\alpha_p$ for the case
$\varepsilon=10^{-3}$ is presented. The curves $1,2,3,4$ correspond
to the following values of the wave number
$k=0.05$, $0.1$, $0.15, 0.25$.

On the Fig. 6  the dependence of the angle $\phi=\arg K$ (of the
phase shift) on the quantity $\varepsilon$ for the case $k=0.2$ is
represented. The curves $1,2,3$ correspond to the following values
of the accommodation coefficient $\alpha_p=0.1,0.5,1$. The analysis
shows that the dependence between the values of the angle $\phi$ and
the wave number and the accommodation coefficient as well is small.

\begin{center}\bf
  9. CONCLUSION
\end{center}

In the present work new boundary conditions for the questions of
plasma wave reflection from the plane boundary of a half--space of
degenerate plasma were proposed. These boundary conditions are
naturally called as specular accommodative conditions. Such boundary
conditions are most adequate for the problems of normal propagation
of plasma waves (perpendicular to the boundary), since accommodation
coefficient under such boundary conditions is normal electron
momentum accommodation coefficient.

In the present paper the analytical solution of the problem of
plasma wave reflection from a boundary with normal electron momentum
accommodation is obtained. The analysis of the main parameters of
the problem in long wave limit is carried out. This analysis shows
that the boundary conditions proposed are intermediate between pure
specular and pure diffuse boundary conditions. Indeed, from the Fig.
3 it is seen that all the graphs showing the dependence between the
reflectance and the wave number are located between graphs
corresponding to specular and diffuse boundary conditions.

\end{document}